\documentclass[prd,onecolumn,nofootinbib,showpacs,showkeys]{revtex4}
\usepackage{graphicx}
\def\bea{\begin{eqnarray}}
\def\eea{\end{eqnarray}}

\def\beq{\begin{equation}}
\def\eeq{\end{equation}}

\newbox\pippobox

\begin{document}
\input epsf
\title{De Sitter nonlinear sigma model and  accelerating universe}
%\author{Soon-Tae Hong} \email{soonhong@ewha.ac.kr}
%\affiliation{Department of Science Education and Research Institute
%for Basic Sciences,\\ Ewha Womans University, Seoul 120-750 Korea}
\author{Joohan Lee}
\email{joohan@kerr.uos.ac.kr} \affiliation{Department of Physics,
University of Seoul, Seoul 130-743 Korea}
\author{Tae Hoon Lee}\email{thlee@ssu.ac.kr}
\affiliation{Department of Physics and Institute of Natural
Sciences,\\ Soongsil University, Seoul 156-743 Korea}
\author{Tae Yoon Moon} \email{dpproject@skku.edu} \author{Phillial Oh}
\email{ploh@dirac.skku.ac.kr}\affiliation{Department of Physics and
Institute of Basic Science, Sungkyunkwan University, Suwon 440-746
Korea}
%\date{January 22, 2003}
\date{\today}
\begin{abstract}
We consider a cosmology with a noncompact nonlinear sigma model. The
target space is of de Sitter type and four scalar fields are
introduced. The potential is absent but cosmological constant term
$\Lambda$ is added. One of the scalar fields is time dependent and
the remaining three fields have no time dependence but only spatial
dependence. We show that a very simple ansatz for the scalar fields
results in the accelerating universe with an exponential expansion
at late times. It is pointed out that the presence of the energy
density and pressure coming from the spatial variation of the three
scalar fields plays an essential role in our analysis which includes
$\Lambda=0$ as a special case and it discriminate from the standard
$\Lambda$-dominated acceleration. We perform a stability analysis of
the solutions and find that some solutions are classically stable
and attractor. We also present a nonperturbative solution which
asymptotically approaches an exponential acceleration and discuss
possible cosmological implications in relation to dark energy.
 It turns out that the
equation of state approaches asymptotically $\omega =-1$ both from
above and below, but the crossing does not occur. It predicts
present value of $\omega\sim -1\mp 0.07$, which is within the region
 allowed by the observational
 data. This solution also exhibits a
power law expansion at early times, and the energy density of the
scalar fields mimics that of the stiff matter.
\end{abstract}
%\pacs{11.30.Pb, 11.30.Qc, 12.60.Jv, 14.80.Hv, 98.80.Cq}
%\preprint{hep-th/yymmnn}
\maketitle

\section{Introduction}
The recent cosmological observations \cite{ob} provide many precise
data and arouse an explosion of recent interests in the cosmology.
The most recent data and its cosmological interpretation
\cite{komatsu} indicate that about $73\%$ of our Universe is made of
dark energy, the origin of which is one of the greatest puzzles in
the modern cosmology \cite{inf}.
% more summary

It is highly conceivable that the dark energy is responsible for the
late acceleration of the Universe \cite{carroll} and many candidates
have been proposed. The simplest approach for the accelerating
universe is to introduce the cosmological constant
\cite{weinberg,carol,sahni,padm,peeb} for the dark energy. Other
approaches \cite{carroll} include dynamical models of the
cosmological constant
\cite{ford,wetterich,quin1,pe,fujii,quin2,cope,ferreira,zlatev,
stein,staros2,kessen,a,m,chiba}. Among them, most commonly proposed
candidates are the quintessence, which is described by a scalar
field minimally coupled to Einstein gravity with a
potential\cite{quin1,quin2}. It is shown that the scalar energy
density is subdominant in the matter dominated, and then,
acceleration takes over at later stage of the cosmological
evolution.
%K-essence model introduces non-canonical kinetic
 Later  the phantom model with a negative kinetic
energy scalar field was proposed \cite{cal1} to account for the
region where the equation of state is less than $\omega=-1$, and
quintom model where the ordinary scalar and the phantom are both
introduced \cite{guo} to explain the crossing of the  $\omega=-1$
line. In these models and subsequent works, various forms
\cite{wetterich,quin1,pe,quin2,ferreira,potential} of the potential
are introduced
% form of
%the potential is introduced
to account for the dark energy and the asymptotic acceleration is
achieved through the quintessence or phantom dominance at late
stage.
%The nature of the acceleration is mostly given by the power
%law acceleration. %For example, in the well-known tracker solution,

%More on motivation
% One of the simple ways to explain it is through
%an introduction of the cosmological constant which leads to an
%exponential expansion of the scale factor of the universe via
%$a(t)\sim e^{\Lambda t}$.

In this paper, we consider a cosmological model with a nonlinear
sigma model \cite{ketov} with a cosmological constant term. The
target space is noncompact and is of four-dimensional de Sitter
manifold and four scalar fields are introduced. One of the
motivations is that the scalar fields have a geometric origin and
the potential term is not necessary. Another motivation is to
consider the spatial dependence of the scalar fields and examine its
consequences. To solve the Einstein equation, we assume that only
one of the scalar fields is time dependent and the remaining three
fields have no time dependence but only spatial dependence. We first
show that a very simple ansatz can solve the Einstein equation which
describes the acceleration of the universe with an exponential
expansion at late stage. The spatial contribution in combination
with the cosmological constant forms an effective cosmological
constant and  plays important roles in order to provide the
necessary energy density and pressure. We will also show that the
acceleration is possible even without the cosmological constant.
 It seems that this feature of
contributing the energy density and pressure coming from the spatial
variation of the scalar fields was not considered before in relation
with dark energy.
%The physical background for phantom type of matter with strongly
%negative pressure may be looked for in string theory.

Then, we perform a stability analysis of the solutions.
 We find that  some of the solutions, depending on the
values of the parameter given, are classically stable and attractor
solutions. We will consider two cases where the target space has
signature $(+,-,-,-)$ or $(-,+,+,+)$. Especially, in the $(+,-,-,-)$
case, the linear stability analysis fails, but we are able to find
out that there exists a nonperturbative solution which
asymptotically approaches the de Sitter acceleration, but at early
times it is a power law expansion. In this case, the cosmological
constant term is uniquely fixed in terms of the other parameter.
These features differentiate the present analysis from the standard
$\Lambda$-dominated late-time exponential acceleration.

One might think that adding a cosmological constant term with the
scalar fields could be {\em ad hoc}, but it seems that at present,
the dynamical models of the dark energy is not completely successful
in solving the cosmological constant problem and many of them
require some kind of fine tuning anyhow. Nevertheless, the aim of
this paper is not to explain the smallness of the cosmological
constant (the fine tuning problem in our approach is mentioned in
the Conclusion and Discussion section), but to focus on the
late-time exponential acceleration of the universe and stability of
its behavior. It turns out that the accelerating universe requires
some bound on the original cosmological constant term. In
$(-,+,+,+)$ case, it require that the original cosmological constant
term must be negative for stability, still the acceleration is
possible and can be led by the scalar fields.
%It combines with
%contribution from the scalar fields to form an effective
%cosmological constant, but fine tuning seems to be inevitable.

The paper is organized as follows. In Sec. II, we present noncompact
nonlinear sigma model coupled with Einstein gravity with a
cosmological constant term and discuss the ansatz which solves the
equations in some generality. In Sec. III, we describe the
exponential accelerating solution with de Sitter target space. In
Sec. IV, the stability analysis is performed and allowed range of
the cosmological constant is classified. In Sec. V, a
nonperturbative solution is obtained and possible cosmological
implications in relation to dark energy is given. Section VI
includes
the conclusion and discussion.\\

\section{The Action and cosmological constant }

 We consider an action in which the Einstein gravity is coupled to a
 nonlinear sigma model with a cosmological constant term (in units of $M_p=1$):
\begin{eqnarray}
S=\int \,d^4x
\sqrt{-g}~[~\frac{1}{2}R-\frac{g^{\mu\nu}}{\lambda^2}G_{\alpha\beta}(\Phi)
\partial_{\mu}\Phi^{\alpha}\partial_{\nu}\Phi^{\beta}-\Lambda+ {\cal L}_{matter}~]
\label{act189}
\end{eqnarray}
where $\Phi^{\alpha}=(\phi,\sigma^{i})~(i=1,2,3)$, $G_{\alpha\beta}$
is the metric of the noncompact target space, $\lambda^2$ is the
self-coupling constant of the nonlinear sigma model and it is
assumed to be positive. $\Lambda$ is the cosmological constant.  The
equations of motion are given by
\begin{eqnarray}
R_{\mu\nu}=\frac{2}{\lambda^2}G_{\alpha\beta}\partial_{\mu}\Phi^{\alpha}
\partial_{\nu}\Phi^{\beta}+\Lambda g_{\mu\nu}+\hat T_{\mu\nu}\label{eom00}\\
\frac{1}{\sqrt{-g}}\partial_{\mu}[\sqrt{-g}g^{\mu\nu}G_{\alpha\beta}\partial_{\nu}\Phi^{\beta}]
=\frac{1}{2}\frac{\partial
G_{\beta\gamma}}{\partial\Phi^{\alpha}}g^{\mu\nu}\partial_{\mu}
\Phi^{\beta}\partial_{\nu}\Phi^{\gamma}, \label{eom2}
\end{eqnarray}
where $\hat T_{\mu\nu}~(= T_{\mu\nu}-g_{\mu\nu}T/2)$ is assumed to
take the perfect fluid form;
\begin{eqnarray}
 T^{\mu}_{\nu}=(-\rho_m,~p_m,~p_m,~p_m).~~
\end{eqnarray}
The matter sector satisfies the continuity equation; $\nabla_{\mu}
T^{\mu\nu}=0.$

 If we ignore the matter part, we can solve the Eqs. (\ref{eom00}) and
 (\ref{eom2}) with the following ansatz
\begin{eqnarray}
\phi=~t,~~\sigma^{i}=x^{i}.\label{sol1}
\end{eqnarray}
To check whether  Eq. (\ref{eom00}) can be solved (without $\hat
T_{\mu\nu}$) with this ansatz, first note that if $\Lambda=0$,
$g_{\mu\nu}(t, x^i)=+ G_{\mu\nu}(\phi,\sigma^{i})$ satisfies the
equation as long as the scalar curvature of the space-time metric
$g_{\mu\nu}$ and that of the target space $G_{\alpha\beta}$ are
constants. Then, we can add an cosmological constant $\Lambda$ which
has the same sign with these curvatures and the equation can still
be satisfied \cite{omero,gellmann}. This has the effect of scaling
the space-time metric via $g_{\mu\nu}\rightarrow
 (1+4\Lambda/R)g_{\mu\nu} $, where $R$ is the scalar curvature constant.
 The cosmological constant
 could even have some value of the opposite sign with these curvatures as long
as the absolute value of the cosmological constant is smaller than
that of the target space, i.e., $\vert \Lambda\vert< \vert R
\vert/4$. This point can be extended further. Suppose $g_{\mu\nu}(t,
x^{i})=- G_{\mu\nu}(\phi,\sigma^{i})$. Then, the scalar curvatures
of the space-time and target space have opposite signature and the
equation cannot be satisfied with $\Lambda=0$. But if we add
$\Lambda$ such that the sign  is the same as the space-time scalar
curvature constant and the absolute value is greater than the scalar
curvature of the target space, the equation can be satisfied. Also,
one can check that  the metric ansatz $g_{\mu\nu}(x)=\pm~
G_{\mu\nu}(\phi,\sigma^{i})$ satisfies the Eq. (\ref{eom2}). In
summary, we find that the metric ansatz and (\ref{sol1}) solve
(\ref{eom00}) and (\ref{eom2}), and we have
\begin{eqnarray}
R_{\mu\nu}=(\pm\frac{2}{\lambda^{2}}+\Lambda)g_{\mu\nu},
\label{sol2}
\end{eqnarray}
as long as the constant curvature condition is satisfied and without
$\hat T_{\mu\nu}$.
% we first find that the solutions are given by

 The $\sigma^{i}=x^{i}$ ansatz first appeared in higher
 dimensional gravity theory in association with spontaneous
 compactification of the extra dimensions \cite{omero, gellmann}.
 It does not break the isotropy and homogeneity of the universe
 as long as we do not introduce the potential for the $\sigma$
 fields.
 Also the
 $\phi=t$  has been exploited
 to unify early-time and late-time universe based on phantom
cosmology \cite{nojiri}. Note that the quantity $\Lambda_{eff}\equiv
\pm 2/ \lambda^{2}+\Lambda$ plays the role of the effective
cosmological constant and there is curvature constant restriction on
the value of $\Lambda$; $\pm 2/ \lambda^{2}+\Lambda$ must have the
same signature as that of the space-time scalar curvature constant.
The above aspect of the ansatz (\ref{sol1}), (\ref{sol2}) is quite a
general feature of the nonlinear sigma model coupled to gravity.
 In this paper, we
will consider the de sitter target space with
\begin{eqnarray}
G_{\alpha\beta}^{(\epsilon)}=\epsilon(~1,~-e^{2\xi\phi},~-e^{2\xi\phi},~-e^{2\xi\phi}~)
\label{de1}
\end{eqnarray}
with $\epsilon=\mp 1$ and $\xi$ being an arbitrary positive
constant. It turns out that the spatial ansatz (\ref{sol1}) provides
contribution of the energy density and pressure such as to reveal
diverse aspects of the late time exponential acceleration, not
present in the standard cosmological constant dominated
acceleration.  We will also find that the allowed value of the
cosmological
constant divides further if required the stability. \\

\section{De Sitter solution}
To discuss the cosmological implication of the solution
(\ref{sol1}), (\ref{sol2}) with the de Sitter target space metric
(\ref{de1}),
 we introduce the standard space-time metric via
\begin{eqnarray}
ds^{2}=-dt^2+a^2(t) dx_{i}dx^{i}.
\end{eqnarray}
With $H=\dot{a}/a$, the equation of motion (\ref{eom00}) becomes
\begin{eqnarray}
H^2&=&\frac{2}{3\lambda^2}~[~\epsilon(\frac{1}{2}\dot{\phi}^2-\frac{1}{2}e^{2\xi\phi}\dot{\sigma^{i}}^2
+\frac{1}{2a^2}(\partial_{i}\phi)^2-\frac{1}{2a^2}e^{2\xi\phi}(\partial_{i}\sigma^{j})^2)+
\frac{\lambda^2}{2}\Lambda]+\frac{1}{3}\rho_{m}\label{mom1}\\
\dot{H}&=&-\frac{1}{\lambda^2}~\epsilon[~\dot{\phi}^2-e^{2\xi\phi}\dot{\sigma^{i}}^2-\frac{1}{3a^2}
(\partial_{i}\phi)^2-\frac{1}{3a^2}e^{2\xi\phi}(\partial_{i}\sigma^{j})^2~]-\frac{1}{2}(1+\omega_{m})\rho_{m},
\label{mom2}
\end{eqnarray}
where $\omega_m=p_m/\rho_m$. The continuity equation implies $\rho_m
\propto a^{-3(1+\omega_m)}.$
 Plugging the ansatz $\sigma^{i}=x^{i}$ and $\phi\equiv \phi(t)$ into the above equations,
 Eqs. (\ref{eom2}), (\ref{mom1}) and (\ref{mom2}) become
 \begin{eqnarray}
 0&=&\ddot{\phi}+3H\dot{\phi}-3\xi\frac{e^{2\xi\phi}}{a^2},\label{h3}\\
H^2&=&\frac{2}{3\lambda^2}~[~\epsilon(\frac{1}{2}\dot{\phi}^2-\frac{3}{2a^2}e^{2\xi\phi})+
\frac{\lambda^2}{2}\Lambda~]+\frac{1}{3}\rho_{m}\label{h1}\\
\dot{H}&=&-\frac{1}{\lambda^2}~\epsilon[~\dot{\phi}^2-\frac{1}{a^2}e^{2\xi\phi}~]
-\frac{1}{2}(1+\omega_{m})\rho_{m}. \label{h2}
\end{eqnarray}
The second terms in both (\ref{h1}) and (\ref{h2}) are the
contributions coming from the spatial variations of $\sigma_i$ which
is essential for the subsequent analysis.
% Note that for matter fields with $\omega_m \geq 0,$ $\rho_m$
% decreases faster than the 2nd term in equation (\ref{h1}) and
% (\ref{h2}) as the universe expands. Moreover,

The scalar dominance
 requires a check of whether the matter contribution term can be
 ignored at late times.
 Our solution corresponds to a linearly increasing scalar field
 with positive $\xi$ such that the kinetic energy terms and the second terms in both (\ref{h1}) and (\ref{h2})
 are constant. Therefore, the contribution of the matter density which decreases as $a^{-3(1+\omega_m)}$ becomes negligible
  at  late times and we ignore
 the matter part here after.
 Now, substitution of $\phi=~t$ leads to
\begin{eqnarray}
a(t)=e^{\xi
t},~~\xi=\sqrt{-\frac{2\epsilon}{3\lambda^2}+\frac{\Lambda}{3}}
~~(\Lambda>\frac{2\epsilon}{\lambda^2}).\label{desitter}
\end{eqnarray}
In the above equation, we fixed the initial values by $a(0)=1$ and
$\phi(0)=0$. Later, we will relax these initial conditions and
accommodate more general conditions. This describes a de Sitter
expansion of the universe. From here on, we will always assume
$\sigma^{i}=x^{i}$ and study the time-dependent behavior of the Eqs.
 (\ref{h3}), (\ref{h1}), (\ref{h2}). Note that for the $\epsilon=+1$
case, the cosmological constant term has to be bigger than some
positive value. In contrast, for the $\epsilon=-1$ case, it could be
any value greater than a fixed negative value and we will see that
stability adds further restrictions. It must especially be negative
for stability; still the universe can accelerate and this is driven
by the scalar fields. The cosmological constant term combines with
the contribution from the scalar fields to form an effective
cosmological constant $\Lambda_{eff}=3\xi^2$ in Eq.
(\ref{desitter}).

\section{Stability}
To check the stability of the above solution, we first consider the
following quantities,
\begin{eqnarray}
2\xi\phi-2N=X,~~N=\ln{(a)}\label{X}
\end{eqnarray}
Plugging (\ref{X}) into (\ref{h3})$\sim$(\ref{h2}), we obtain
\begin{eqnarray}
3H^2+\dot{H}=-\frac{2\epsilon}{\lambda^2}e^{X}+\Lambda\label{h4}\\
\nonumber\\
\ddot{X}+3H\dot{X}-(6\xi^2+\frac{4\epsilon}{\lambda^2})e^{X}+2\Lambda=0\label{h5}
\end{eqnarray}
 The solution (\ref{desitter}) corresponds to $X=0$ with
$ H=\xi, ~\dot{\phi}=1$. In order to accommodate more initial
conditions, we consider the solution $X=X(0)\equiv \ln f$.  Then,
the effective cosmological constant becomes
$\Lambda_{eff}(f)=-2\epsilon f/\lambda^{2}+\Lambda$, and the
solution (\ref{desitter}) is replaced by
\begin{eqnarray}
\phi=~\sqrt{f}t+\phi(0),~a(t)=a(0) e^{\sqrt{f}\xi
t},~~\xi=\sqrt{-\frac{2\epsilon}{3\lambda^2}+\frac{\Lambda}{3f}}
,\label{desitter1}
\end{eqnarray}
with $2\xi \phi(0)-2\ln a(0)= \ln f,~\Lambda>\frac{2\epsilon
f}{\lambda^2}$. Note that the exponent $\xi$ behaves under the
change of the initial conditions when $\Lambda=0$ as follows; $f
\rightarrow g, e^{\sqrt{f}\xi t}\rightarrow e^{\sqrt{g}\xi t}$.

 The linear perturbation of Eq. (\ref{h5}) leads to
\begin{eqnarray}
\delta\ddot{X}+3\sqrt{f}\xi\delta\dot{X}-(6\xi^2+\frac{4\epsilon}{\lambda^2})f\delta
X=0\label{XX}
\end{eqnarray}
Introducing $\delta X  \sim e^{\gamma t}$, Eq. (\ref{XX}) yields
\begin{eqnarray}
\gamma^2+3\sqrt{f}\xi\gamma-(6\xi^2+\frac{4\epsilon}{\lambda^2})f=0
\label{gamma}
\end{eqnarray}
The solutions for the Eq. (\ref{gamma}) are
\begin{eqnarray}
\gamma_{+}/\sqrt{f}&=&\frac{-3\xi+\sqrt{33\xi^2+\frac{16\epsilon}{\lambda^2}}}{2}\label{root1}\\
\gamma_{-}/\sqrt{f}&=&\frac{-3\xi-\sqrt{33\xi^2+\frac{16\epsilon}{\lambda^2}}}{2}\label{root2}
\end{eqnarray}
%For $\epsilon=+1$ case, $\gamma_{+}>0$ and $\gamma_{-}<0$ for any
%$\xi$. Hence, the solution is unstable. It corresponds to a saddle
%point. \\
%For $\epsilon=-1$ case, there are two cases. In the case
%$\frac{16}{33\lambda^2}<\xi^2< \frac{2}{3\lambda^2}$, $\gamma_{\pm}$
%are both negative. And the solution is stable. For
%$\xi^2>\frac{2}{3\lambda^2}$, $\gamma_{+}>0$ and $\gamma_{-}<0$, so
%the solution is unstable. In the case
From these equations, we have the following cases:\\
 (A). For
the $\epsilon=+1$ case, $\gamma_{+}>0$ and $\gamma_{-}<0$ for any
$\xi$. Hence, the solution is unstable. It corresponds to a saddle
point. The cosmological constant has to be positive.
\\
For the $\epsilon=-1$ case, there are three cases. In all three
cases, the cosmological constant has to be negative for stability,
but the late
acceleration is achieved by the scalar fields.\\
 (B). In the case
$\frac{16}{33\lambda^2} \leq \xi^2< \frac{2}{3\lambda^2}$,
$\gamma_{\pm}$ are both negative and the solution is stable and an
attractor. For $\xi^2=\frac{16}{33\lambda^2}$, the root is
degenerate with $\gamma_+/\sqrt f=\gamma_-/\sqrt f=-3\xi/2$.
\\
 (C). For
$\xi^2>\frac{2}{3\lambda^2}$, $\gamma_{+}>0$ and $\gamma_{-}<0$, so
the solution is unstable and corresponds to a saddle point.\\
 (D). In the case
$\xi^2<\frac{16}{33\lambda^2}$, $\gamma_{\pm}$ becomes imaginary and
the perturbation is oscillatory and it is an attractor.
% $\xi^2<\frac{16}{33\lambda^2}$,
%$\gamma_{\pm}$ becomes imaginary and the perturbation is oscillatory.

The linear perturbation can be integrated explicitly. For (A), (B)
and (C), Eq. (\ref{XX}) yields
\begin{eqnarray}
\delta X&=&Ae^{\gamma_{+}t}+Be^{\gamma_{-}t}\\
\delta H&=&\frac{-2\epsilon f}{\lambda^2}(\frac{A
e^{\gamma_{+}t}}{\gamma_{+}+6\sqrt{f}\xi}+ \frac{B
e^{\gamma_{-}t}}{\gamma_{-}+6\sqrt{f}\xi}-C_{0}e^{-6\sqrt{f}\xi
t}),\label{singular}
\end{eqnarray}
where
$C_{0}=A/(\gamma_{+}+6\sqrt{f}\xi)+B/(\gamma_{-}+6\sqrt{f}\xi)$.
 Note that when $\epsilon=+1$, the second term in Eq.
 (\ref{singular}) diverges for the value $\xi^2=1/3\lambda^2$.
 This might imply that linear perturbation fails in this case and in
 fact, there exist a nonperturbative solution as will be discussed
 in the next section. For case (D), we have
%\begin{eqnarray}
%\gamma_{+}&=&\frac{-3\xi+\sqrt{33\xi^2-\frac{16}{\lambda^2}}}{2}>0\\
%\gamma_{-}&=&\frac{-3\xi-\sqrt{33\xi^2-\frac{16}{\lambda^2}}}{2}<0
%\end{eqnarray}
\begin{eqnarray}
\delta X=D\cos(\omega t+\theta_{0})e^{-\frac{3}{2}\sqrt{f}\xi t}
\end{eqnarray}
where $\omega^2=\frac{f}{4}(\frac{16}{\lambda^2}-33\xi^2)$ and
\begin{eqnarray}
\delta H=-E e^{-6\sqrt{f}\xi t}+F(t) e^{-\frac{3}{2}\sqrt{f}\xi t}
\end{eqnarray}
where \begin{eqnarray} E&=&36Df\sqrt{f}\xi\cos{\theta_{0}}
/\lambda^2(81f\xi^2+4\omega^2)+8Df\omega\sin{\theta_{0}}/\lambda^2(81f\xi^2+4\omega^2)
 ,\nonumber\\
~F(t)&=&36Df\sqrt{f}\xi\cos{(\omega t+\theta_{0})}
/\lambda^2(81f\xi^2+4\omega^2)+8Df\omega\sin{(\omega
t+\theta_{0})}/\lambda^2(81f\xi^2+4\omega^2).
\end{eqnarray}
% This is an attractor solution.

We comment on the case $\xi^2 = 2/3\lambda^2$ separately. It
corresponds to when the cosmological constant $\Lambda$ is zero. In
this case, one of the roots $\gamma_{+}$ of Eq. (\ref{root1})
becomes zero, and the other root $\gamma_{-}$ is negative. Its
stability is indecisive at this level. These solutions, their
stability and contents of the cosmological constant are summarized
in Table I. It is interesting to note that in the $\epsilon=-1$
case, any value greater than a fixed negative of the cosmological
constant was allowed in (\ref{desitter}), but stability requires its
upper bound must be zero.

%%%%%%%%%%%%%%%%%%%%%%%%%%%%%
\begin{table*}[t]
\begin{center}
\begin{tabular}{|c|c|c|c|c|}
\hline \hline
Name &  $\epsilon$ & $\xi$ & Stability & $\Lambda$ \\
\hline \hline (A) & +1 & $\xi > 0$ & $\gamma_{+} >
 0,~\gamma_{-} < 0$, Unstable, saddle point& $\Lambda > 2f/ \lambda^2$  \\
 \hline (B) & $-1$ & 16/33$\lambda^2 \leq \xi^2 < 2/3\lambda^2$&
  $\gamma_{\pm}< 0$ , Stable, attractor & $-6f/11\lambda^2 \leq \Lambda < 0$ \\
\hline (C) & $-1$ & $\xi^2 > 2/3\lambda^2$ & $\gamma_{+}>
0,~\gamma_{-} < 0,$
Unstable, saddle point & $\Lambda > 0$ \\
 \hline (D) & $-1$ & $\xi^2 < 16/33\lambda^2$ &  $\gamma_{\pm}$, Imaginary, stable, attractor &
 $ - 2f/ \lambda^2 < \Lambda < -6f/11\lambda^2$ \\
\hline (E) & $-1$ & $\xi^2=2/3\lambda^2$ &$\gamma_{+}=0,~\gamma_{-}<0$  & $ \Lambda = 0 $ \\
\hline (F) & $+1$ & $\xi^2=1/3\lambda^2$ & Nonperturbative & $\Lambda = 3f/\lambda^2 $ \\
& & & $\phi=\sqrt{f}t+\phi_{0}+A\ln{(1+Ce^{-\sqrt{f}\gamma t})}$ & \\
\hline

\end{tabular}
\end{center}
\caption[crit]{Various accelerating solutions and their stability}
\end{table*}
%%%%%%%%%%%%%%%%%%%%%%%%%%%%%%

% If $X=X_0$ is not equal to zero, but still with $ H=\xi, ~\dot{\phi}=1$, it corresponds to different initial
%values of $\phi$ or the scale factor $a(t)$. The stability
%conditions are unchanged, but  $\xi$ in equation (\ref{desitter}) is
%replaced with

\section{Nonperturbative solution}
It turns out that in the $\epsilon=+1$ case, an explicit
nonperturbative solution can be found. To see that, let us first
assume
 $e^{\xi\phi}=\sqrt{f}a.$ Then, Eq. (\ref{h1}) suggests that
 for nontrivial solution, we must have
\begin{eqnarray}
H^2=\frac{1}{3\lambda^2}\dot{\phi}^2,~~\lambda^2
\Lambda=3f\label{dominant}
\end{eqnarray}
 The first of the above equation yields $\xi^2=1/3\lambda^2$. This value of $\xi^2$
was the one where linear perturbation failed in the previous
section.
 Substituting the
 ansatz $a=\frac{1}{\sqrt{f}}e^{\xi\phi}$ into (\ref{h3}), we obtain
\begin{eqnarray}
\ddot{\phi}+3\xi\dot{\phi}^2-3f\xi=0 \label{phi}
\end{eqnarray}
 Note that this equation describes particle motion where constant external
 force and velocity square dependent frictional force are acting.
When $\dot\phi(0)^2 <f$, the constant force term dictates the
particle motion at early times and it accelerates until the velocity
reaches the terminal velocity $\dot\phi(\infty)=\sqrt{f}.$ When
$\dot\phi(0)^2 >f$, the friction term dominates at early times and
it decelerates until the velocity reaches the terminal velocity
$\dot\phi(\infty)=\sqrt{f}.$

 We can find the solution for the above Eq. (\ref{phi}) as follows
\begin{eqnarray}
\phi(t)&=&\sqrt{f}t+\phi(0)-\frac{1}{3\xi}\ln(1+C)+\frac{1}{3\xi}\ln(1+Ce^{-6\sqrt{f}\xi
t}),\label{phi1}
\end{eqnarray}
with $\xi=\sqrt{\frac{1}{3\lambda^2}}$
 and
\begin{eqnarray}
a(t)&=&a(0) e^{\sqrt{f}\xi t}(\frac{1+Ce^{-6\sqrt{f} \xi
t}}{1+C})^{\frac{1}{3}}.\label{nsol1}
\end{eqnarray}
 The constant C remains arbitrary as long as the validity of the solution
 is confined within  the region  $1+Ce^{-\sqrt{f}\gamma
t}>0$. Note that the solution (\ref{phi1}) has two arbitrary
integration constants and it is a complete solution for the ansatz
 $e^{\xi\phi}=\sqrt{f}a$ with $\phi(0)$ and $\dot\phi(0)= \sqrt{f}(1-2C)/(1+C).$
 %Let us first discuss the solution $\xi=+\sqrt{\frac{1}{3\lambda^2}}$.
 %In this case,
 It indicates that
 starting with an arbitrary value of C, the solution converges
 rapidly to $\phi(t)=\sqrt{f}t$ and $a(t)=e^{\sqrt{f}\xi t}$.
There is a wide range of initial conditions in which the solution
rapidly converges to de Sitter  acceleration.
 %Such a fine tuning inevitably occurs in many existing
 %model of dark energy.

To see the solution more closely, let us divide the case with $C>0$
and $C<0.$ First, note that the condition  $1+Ce^{-\sqrt{f}\gamma
t}>0$ puts a restriction on the range of $\vert C\vert$, i.e.,
$\vert C\vert <1$ when  $C<0.$ When $C>0$, it could be arbitrary. We
introduce a time scale defined by $\vert C\vert =e^{-6 \sqrt{f}\xi
t_*}$. Then, Eq. (\ref{phi1}) can be written as
\begin{eqnarray}
\phi(t)=\left\{\begin{array}{ll}
\frac{1}{3\xi}\ln(\sinh(3\sqrt{f}\xi(t+t_{*})))+\tilde{\phi}(0),
~~(C<0,~0<t_{*}<\infty) \label{soo1}\\
\frac{1}{3\xi}\ln(\cosh(3\sqrt{f}\xi(t+t_{*})))+\tilde{\phi}(0),~~(C>0,~-\infty
<t_{*}<\infty) \label{soo2}\end{array}\right.
\end{eqnarray}
where $\tilde{\phi}(0)=\phi(0)-\ln(\sinh(3\sqrt{f}\xi t_{*}))/3\xi$
for $C<0$ and $\tilde{\phi}(0)=\phi(0)-\ln(\cosh(3\sqrt{f}\xi
t_{*}))/3\xi$ for $C>0$ in (\ref{soo2}). And we also have
\begin{eqnarray}
a(t)=\left\{\begin{array}{ll}
\tilde{a}(0)(\sinh(3\sqrt{f}\xi(t+t_{*})))^{\frac{1}{3}},~~(C<0,~0<t_{*}<\infty)\label{soo3}\\
\tilde{a}(0)(\cosh(3\sqrt{f}\xi(t+t_{*})))^{\frac{1}{3}},~~(C>0,~-\infty
<t_{*}<\infty)\label{soo4}\end{array}\right.
\end{eqnarray}
where $\tilde{a}(0)=a(0)/(\sinh(3\sqrt{f}\xi t_{*}))^{\frac{1}{3}}$
for $C<0$ and $\tilde{a}(0)=a(0)/(\cosh(3\sqrt{f}\xi
t_{*}))^{\frac{1}{3}}$ for $C>0$ in (\ref{soo4}).
 Note that $\dot\phi\vert_{t=-t_*}$ becomes
singular in (\ref{soo1}), but our initial time is chosen to be $0$,
and it is outside the range of dynamics. Had we chosen our initial
time to be $t_i$, the singularity would occur at $t=t_i-t_*.$ We
will assume that $\vert C\vert\sim 1$ so that $t_*$ is nearly the
initial time.

Let us assume that the initial time $t=0$ is chosen when  the
universe is still at the matter-dominated epoch and examine the
early-time behavior of the solutions (\ref{soo1}) and (\ref{soo4}).
 Then, for $C<0$ with Eqs (\ref{soo1}) and (\ref{soo3}), we
 have
 \begin{eqnarray}
\phi(t)\sim \frac{1}{3\xi}\ln (t+t_*) + \tilde\phi_0,~ a(t)\sim
(t+t_*)^{\frac{1}{3}}.
\end{eqnarray}
The logarithmic time-dependent $\phi$ field
\cite{wetterich,quin1,ferreira,cope} also appears in the
quintessence with exponential potential.
%Note that the power law
%in this case corresponds to the equation of state $\omega_\phi=1$,
%and it could be thought of as dark matter candidate.
The energy density $\rho_\phi=\frac{1}{\lambda^2}\dot\phi^2\sim
(t+t_*)^{-2}\sim a^{-6}$ scales the same as the stiff matter density
and is known as a scaling solution. In our case, the scaling
behavior holds only at early times, and as time goes by, the full
solution (\ref{soo1}) will take over.
 For $C>0$ with Eqs. (\ref{soo2})
and (\ref{soo4}), we see that the $\phi$ field and the scale factor
$a(t)$ remains constant up to first order in time, and as time goes
by, both quantities begins to grow. There is no scaling behavior in
this case. Note that for both cases, $\Omega_\phi=
\frac{\rho_\phi}{\rho_c}=1$ due to the first condition of Eq.
(\ref{dominant}). This is because in the above solution, matter
contribution was neglected. It would be interesting to check whether
the dominance of the energy density of scalar fields emerges from
the matter-dominated epoch when the matter contributions are
included.

 %\begin{eqnarray}
%\dot \phi(t)\sim \frac{1}{3\xi}\ln t,~ a(t)\sim constant.
%\end{eqnarray}
%The power law in this case corresponds to the equation of state
%$\omega_\phi\rightarrow \infty$, and the energy density
%$\rho_\phi=\frac{1}{2\lambda^2}\dot\phi^2=$

Let us discuss some issues of dark energy with this solution. First,
we have the acceleration  given by
\begin{eqnarray}
\frac{\ddot{a}}{a}=f\xi^2\frac{1 \pm 10e^{-6\sqrt{f}\xi
(t+t_{*})}+e^{-12\sqrt{f}\xi (t+t_{*})}}{(1\pm e^{-6\sqrt{f}\xi
(t+t_{*})})^2},
\end{eqnarray}
where $+ $ in the numerator is for $C>0$ and $-$ is for $C<0$. We
see that for $C>0$, it is always accelerating. For $C<0$, there is a
transition time where the scale factor changes from deceleration to
acceleration. It is given with $t_*\sim 0$ by
\begin{eqnarray}
t_{tr}=\frac{1}{6\sqrt{f}\xi}\ln(5+2\sqrt{6})\sim 0.48
t_0.\label{trans}
\end{eqnarray}
where $t_0$ is the current age of the universe \cite{ishida}. This
fix $\sqrt{f}\xi\sim 0.79/t_0.$
 Next, equation of state with this value of $\sqrt{f}\xi$ is given by
 \begin{eqnarray}
 \omega=-1\mp 8\frac{e^{-6\sqrt{f}\xi (t+t_{*})}}{(1\mp e^{-6\sqrt{f}
  \xi (t+t_{*})})^2}\sim -1\mp 0.07.\label{omegaapp}
 \end{eqnarray}
 For $C>0$ with an upper sign in (\ref{omegaapp}), $\omega$ approaches $-1$ from below.  $\omega$
 is singular at $t=-t_{*}$ and again it is outside the range of dynamics.
  For $C<0$, it approaches $-1$ from above.
 And there is no singularity.
 In both cases, there is no crossing the $\omega=-1$, but
 asymptotically approaches the $\omega=-1$ line.
 We find that this number for the equation of
 state in  (\ref{omegaapp}) is in the region
 allowed by the observational
 data \cite{komatsu}, but inclusion of matter might change the
 result somewhat.
% Keeping the matter contribution would not change the stability of
 %the above solution at late times because the matter energy density decreases
 %proportional to $1/a^3(t)$ and decays rapidly.

\section{Conclusion and Discussion}

 We presented a de Sitter nonlinear sigma model coupled to Einstein gravity in order to
 describe the current acceleration
 of the Universe. It has some characteristic features as follows.
Out of the four scalar fields, only one of them is time-dependent
and the remaining three fields have only spatial dependence. If the
time dependent scalar field is phantom, then the remaining fields
are ordinary scalar fields, or vice versa. The formal case could
also be thought of as the dilatonic phantom coupled with triplet of
scalar fields, whereas the latter case  as the dilaton coupled with
triplet of phantom scalar fields. Since kinetic energy of both
positive and negative sign exists, it could be thought of as a
quintom model \cite{guo} with dilaton interaction between the two
sectors. But the quintom model only considers  time-dependent
fields.  A specific form of the potential is not needed to achieve
the late-time exponential acceleration, but introduction of a
potential could produce subdominant behavior of the scalar fields as
in the quintessence. It is suspected that the potential does not
modify the late-time exponential behavior, because it does not
change the Eq. (\ref{h2}).

We find that a simple ansatz provides the constant energy density
and  results in an accelerating universe with an exponential
expansion. The balance between the pressures coming from the
time-dependent field and spatial-dependent fields makes it possible
to achieve the exponential acceleration. It is pointed out that the
target space of Euclidean de Sitter space with signature $(+,+,+,+)$
cannot produce such balance and exponential acceleration of the
universe. The model has essentially two parameters, $\xi$ and the
cosmological constant term $\Lambda$. $\sqrt{f}\xi M_p$ plays the
role of the Hubble constant and is a function of the strength of the
self-coupling constant and the cosmological constant term $\Lambda$.
Consider, for example, the nonperturbative case with
$\xi^2=1/3\lambda^2$ and $\Lambda=3f/\lambda^2$. Recall
$f=e^{2\xi\phi(0)/M_p}/a^2(0)$ and let us assume $a(0)\sim 1$. We
mention a couple of cases where $\sqrt{f}\xi M_p\sim 10^{-61} M_p$
and $\Lambda\sim 10^{-122}M_p^2$ can be realized. In the first case
with $\phi(0)\sim -M_p$ and $\lambda\sim 1/250$, we have $\xi
\phi(0)\sim -145 M_p$.
 If the scale
when the nonlinear sigma model sets in is of the order of Gev with
$\phi(0)\sim -1 GeV$, this requires extremely weak coupling constant
with $\lambda\sim 10^{-21}$. In this case, we have  $\xi \phi(0)\sim
-190 M_p.$ These are fine tunings which can yield the small Hubble
constant and the cosmological constant.

 Stability
analysis shows some of the solutions, depending on the values of the
parameter $\xi$, are classically stable and attractor solutions.
They require that the original cosmological constant term must be
negative, still the acceleration is possible led by the scalar
fields.
 In one case, where the cosmological constant term is
uniquely fixed, there is a nonperturbative solution which
asymptotically approaches the de Sitter phase of acceleration. This
solution also exhibits a power law expansion at early times, and the
energy density of the scalar fields mimics the matter energy
density. It remains to be seen whether the stability survives when
the analysis is extended to spatial variations.

%These features differentiate the present analysis from the standard
%cosmological constant dominated late time acceleration.

The present analysis indicates that the acceleration phase can be
dominated by the nonlinear sigma  model. We only focused on the late
time behavior except the nonperturbative case. To show whether this
behavior of scalar dominance can emerge from matter-dominated epoch,
the analysis has to be extended including the contribution of matter
density at early times which was neglected.
%One of the remaining issues
%is that the existence of the scaling behavior of the energy density
%\cite{..} of the present model has to be checked. In order to to
%that,
Finally, whether the de Sitter nonlinear sigma model could come from
particle physics as an effective low energy field theory remains to
be seen. These aspects needs further investigation.\\

 {\bf Note
added}: After the completion of this work, we became aware of Ref.
\cite{dubo} where the ansatz (\ref{sol1}) each multiplied by some
constant factors to have Minkowski background also appeared in the
cosmological context of the Lorentz violating massive graviton
models \cite{hooft}.  These models deal with flat background metric.
However, in our de Sitter background solution, a linear perturbation
of the metric in the action (\ref{act189}) does not result in any
massive graviton mode even though the ansatz (\ref{sol1})
spontaneously break the diffeomorphism invariance. This can be
readily seen by checking that the mass term which is of the second
order in the perturbations cancels out in the second term of the
action in (\ref{act189}). Moreover, the de Sitter background
solution allows the modification where each of the ansatz in
(\ref{sol1}) can be multiplied by a {\it same} constant factor only
which still does not yield Lorentz violating mass term. Perhaps, it
could be possible to generate a mass term by a suitable deformation
of the target space metric in the action (\ref{act189}), which is
beyond the scope of this work, but nevertheless whose implications
would be worthwhile to be explored in detail.

\section{Acknowledgement}
We  thank the anonymous referee for pointing out Ref. \cite{dubo} to
us.  We also thank S. T. Hong for early participation and Y.-Y. Keum
for useful information on dark energy.  The work of THL was
supported by the Korea Research Foundation Grant
 funded by the Korean Government (MOEHRD, Basic Research Promotion
 Fund) (KRF-2007-313-C00165).
The work of PO  is supported by the Science Research Center Program
of the Korea Science and Engineering Foundation through the Center
for Quantum Spacetime(CQUeST) of Sogang University with Grant No.
R11-2005-021.

\end{document}